\documentclass[twocolumn,preprintnumbers,amsmath,amssymb]{revtex4}

\usepackage{graphicx}
\usepackage{dcolumn}
\usepackage{bm}
\usepackage{epsfig} 

\begin{document}


\title{Effect of rolling on dissipation in fault gouge}



\author{F. Alonso-Marroqu\'{\i}n}
\email{fernando@esscc.uq.edu.au}
\affiliation{ ESSCC, The University of Queensland, Qld. 4068, Brisbane, Australia}
\affiliation{  National Technical University of Athens, 5 Heroes of
              Polytechnion, 15773, Athens, Greece}

\author{I. Vardoulakis }
\affiliation{ National Technical University of Athens, 5 Heroes of
              Polytechnion, 15773, Athens, Greece}

\author{H. J. Herrmann}
\affiliation{ ICP, University of Stuttgart, Pfaffenwaldring 27, 
              70569, Stuttgart, Germany}
\affiliation{ IfB, HIF E.11, ETH H\"onggerberg CH 8093 Z\"urich, Switzerland  }

\author {D. Weatherley and P. Mora}
\affiliation{ESSCC, The University of Queensland, Qld. 4068, Brisbane, Australia }

\date{\today}

\begin{abstract}
Sliding and rolling are two outstanding deformation modes in granular 
media. The first one induces frictional dissipation whereas the latter one 
involves deformation with negligible resistance. Using numerical simulations 
on two-dimensional shear cells, we investigate the  effect of the grain 
rotation on the energy dissipation and the strength of granular materials 
under  quasistatic shear deformation. 
Rolling and sliding are quantified in terms  of the so-called 
{\it Cosserat rotations}. The observed spontaneous formation 
of vorticity cells and  clusters of rotating bearings may provide an 
explanation for the long standing heat flow paradox of earthquake
dynamics.

\end{abstract}


\maketitle


\section{Introduction}

The micromechanics of heat production by friction in granular materials 
has become an important issue in the study of earthquake mechanics. One of 
the unresolved controversies in this field is a phenomenon that geophysicists 
call the heat-flow paradox \cite{mora99}. According to common sense, 
when two blocks grind against one another, there should be friction, 
and that should produce heat. 
However, measurements of heat flow during earthquakes are unable to 
detect the amount of heat predicted by simple frictional models. Calculations 
using the value of rock friction measured in the laboratory, i.e. a typical 
friction coefficient between $0.6$ and $0.9$ \cite{byerlee78}, 
lead to overestimation of the heat flux. 
As an example one refers in this context to the heat flow observations 
made around the San Andreas fault, which show that the effective friction 
coefficient must be around $0.2$ or even less~\cite{lachenbruch92}. 
One possible scenario for the 
explanation of these observations is the mechanism of heat-induced pore-fluid 
pressure increase \cite{lachenbruch92,sulem04,rise06}. Other mechanisms have 
been  also discussed \cite{mora99,mahmoodi04,wilson05}.  
At any rate,  with or without pressurization, the correct 
assessment of frictional heat production during shear is a central issue. 
Here we will address this issue by resorting to the micromechanics of dry 
granular media representing the gouge, i. e. the shear band 
consisting of fragmented rock inside the fault zone.

The formation of rolling bearings inside the gouge has 
been introduced as a possible explanation for a substantial reduction of the 
effective friction coefficient \cite{mahmoodi04}.
This simplified picture assumes that the gouge is filled with more of less 
round grains which, as the plates move, can roll on each other thus reducing 
the amount of frictional dissipation. Granular dynamics simulations 
\cite{mora99,zervos00, astrom00} and Couette experiments \cite{veje99}  
have demonstrated the spontaneous formation of such bearings.

The overall effects of grain rotation are studied here using granular
dynamics simulations.
We show that particle rotation induces a phase separation in the granular 
media in terms of three coexisting phases: (1) Vorticity  cells, 
(2) rotational bearings and (3) slip bands. The first two phases reduce
significantly the frictional strength and the dissipation 
with respect to the hypothetical case of simple shear. We quantify 
these phases in terms of the so-called {\it Cosserat rotations}. 
We address the necessity to introduce these variables for the constitutive 
modeling of fault gouge.  In  Secs~\ref{cosserat} and \ref{model} we present
the theoretical background and the particle-based model. In  
Sec.~\ref{rotation} the effect of particle rotation on the strength and 
frictional dissipation is investigated.  In Sec.~\ref{rolling} we calculate 
the population  of the three coexisting phases using the homothetic-antithetic 
decomposition of the Cosserat rotations.

\section{Cosserat Continuum}
\label{cosserat}

In the framework of continuum mechanics, the mathematical description of 
granular rolling and sliding is a challenging task. We notice first that 
classical continuum theories introduce the concept of the material point as a 
representative assembly of grains and ascribes to it only the degrees of 
freedom of displacement, which in turn are correlated to the displacements 
of the grains of the assembly. Thus classical continuum theory makes no 
provision for particle rotation. More recent continuum models include the 
rotational degrees of freedom by using the so-called Cosserat rotations 
~\cite{ehlers03,vardoulakis95b,froiio05a,muhlhaus87b}. 
These are continuum field variables, measuring the relative particle rotations 
with respect to the rotational part of the displacement-gradient field. The 
name of these variables gives tribute to the brothers Cosserat (1909) who were 
the first to propose such a continuum theory. 50 Years after the first
publication of the original work [10], the basic 
kinematics and static concepts of Cosserat  continuum were reworked in a 
milestone paper by G\"unther~\cite{guenther58}. G\"unther's paper marks the 
rebirth of micromechanics in the 1960s. Following this publication, several 
hundred of papers were published on the subject of micromechanics of granular 
media~\cite{satake67}. The growing interest in the Cosserat theories in recent 
years followed the link that was made by M\"uhlhaus and Vardoulakis
\cite{muhlhaus87b}   between  the Cosserat continuum and the onset of 
shear-bands in granular materials. Observations of particle rotations from 
particle-based and continuum-based computer models \cite{papanastasiou89} 
and physical experiments \cite{lerat96} show that inside shear- and 
interface-bands the particles rotate differently as their neighborhood. 
These findings demonstrated the necessity to introduce the 
Cosserat rotations as additional field variables in the shear-band evolution. 
These new variables involved also a characteristic material length that allows 
in  turn  to reproduce the characteristic width of shear bands. This internal
length has special significance from the computational point of view, because 
its resolves the mesh-dependency problems in the Finite 
Element simulations \cite{papanastasiou92,tejchman96}.

At any material point of the Cosserat continuum we assign a velocity  
$\vec v$  and a spin vector $\vec\omega$ . 
Accordingly in plane-strain deformation the kinematic fields are: 
the classical strain-rate tensor, which corresponds  to the symmetric part 
of the average particle velocity :

\begin{equation}
D_{ij} = \frac{1}{2}
(\frac{\partial v_i}{\partial x_j} +\frac{\partial v_j}{\partial x_i}), 
\end{equation}

\noindent
and the Cosserat rotation, that is given by the difference between
the macro- and micro-rotations:

\begin{eqnarray}
W = \frac{1}{2}
(\frac{\partial u_x}{\partial y} -\frac{\partial u_y}{\partial x})-\omega, 
\end{eqnarray}

\noindent
This variable can be calculated as the difference between the rotation 
of the branch  vector - defined as the line connecting particle mass centers
-  and the  rotation of the particle, averaged over all pair contacts in the 
representative volume element.

\section{Particle-Based Model}
\label{model}

\begin{figure}[b]
  \begin{center}
    \epsfig{file=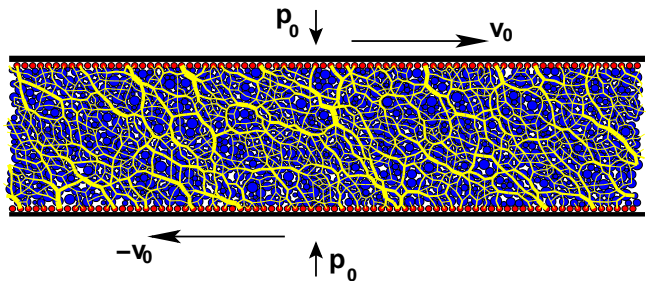,width=\linewidth}
   \caption{Contact network in the shear cell at critical state. The lines 
represent the branch vectors; the width of the lines the normal contact force.}
   \label{fig:cell}
  \end{center}
\end{figure}

We will investigate the discrete counterpart of the Cosserat rotations
by using granular dynamics simulations.
The discrete model consists of disks confined between two horizontal plates. 
A normal  force is  applied on the plates,  as they are moved in opposite  
directions with a constant velocity. 
Periodic boundary conditions are imposed along the
horizontal direction, see Fig.~\ref{fig:cell}.
This  geometry is a simplified model of a gouge. 
Real rock gouge consists of non-spherical particles.  
In the present model we are able to simulate two extreme cases: 
The  first one represents a young fault, which is characterized by
a strong interlocking of closely packed rocks in the gouge. This case
is  simulated by hindering the rotation of the disks. 
In the second case we allow the particles to rotate without rolling 
resistance. This is an idealization of mature gouges, where the 
interlocking and the rock asperities are reduced due to grain fragmentation. 

The discrete model is  a 2D implementation of the Lattice Solid Model 
\cite{mora94,mora99,latham05}.  This  is a suitable platform to investigate  
the dynamics of fault gouge via granular dynamics simulations. 
Two disks of radii $R_i$ and $R_j$ interact when the distance between 
their centers of mass $r_{ij}$ is less that the sum of their radii. 
Their interaction is given by the viscoelastic contact force

\begin{equation}
\vec{f}^c= k_n \Delta x_n\vec{n}+k_t \Delta x_t \vec{t}+
     \delta_0 m(\gamma_n v^c_n \vec{n} + \gamma_t v^c_t \vec{t})
\label{eq:contact force}
\end{equation}

\noindent
The first two terms are elastic forces, and the last
two are viscous forces. 
The unit normal vector $\vec n$ points in the direction 
of the vector connecting the center of mass of the two disks. The  tangential 
vector  $\vec t$ is taken perpendicular to $\vec n$. The elastic 
material constants are the normal $k_n$ and tangential $k_t$ 
grain contact stiffnesses, 
The normal elastic deformation is the overlapping length 
$\Delta x_n =R_i + R_j -r_{ij}$. The tangential 
elastic deformation is chosen to be consistent with the Coulomb 
sliding condition as follows: When two particles come into contact we set 
$\Delta x_t =0$.  Then, at each time $t$, we guess a new value for the  
tangential elastic deformation as:  

\begin{equation}
\Delta x^p_t(t)= \Delta x_t (t-dt) + v^c_t dt,
\end{equation}

\noindent
where $v^c_t$ is the tangential relative velocity at the contact:

\begin{equation}
\label{eq:reltanvel}
v^c_t=\omega_i R_i + \omega_j R_j +(\vec v_i-\vec v_j)\cdot \vec t.
\end{equation}
\noindent

\noindent
Here $\vec{v}_i$ is the velocity and  $\omega_i$ is the  
angular velocity  of the particles in contact. 
The predicted value of elastic deformation should be corrected to
satisfy  the Coulomb sliding condition $|f^e_t|<\mu f^e_n$,
where $f^e_t$ and $f^e_n$ are the tangential and normal 
component of the elastic force and $\mu$ is the friction
coefficient. This condition equates to:

\begin{equation}
\label{eq:dxt}
\Delta x_t (t) = sign(\Delta x^p_t(t)) \min(\frac{\mu k_n \Delta x_n (t)}{k_t},|\Delta x^p_t|).
\end{equation} 

\noindent
The viscous force in  Eq. (\ref{eq:contact force}) assures the restitution 
between colliding particles. $\gamma_n$ and $\gamma_t$ are
coefficients of viscosity, and the harmonic mean  
$m=(1/m_i+1/m_j)^{-1}$ is the effective
mass of the disks.   The normal and tangential components of the relative
velocity at the contacts are $v^c_n=(\vec v_i-\vec v_j)\cdot \vec n$
and $v^c_t$ given by Eq.~(\ref{eq:reltanvel}). The factor
$\delta_0=\Delta x_n/R_0$ is included in Eq.~\ref{eq:contact force}, to 
guarantee continuity of the viscous force during collision. $R_0$ is the 
averaged radius of the disks.

The roughness of the driving plates is modeled by attaching particles
to it. Their vertical displacement is controlled with a simple 
viscoelastic force $f^b=k_n \delta+\gamma_n m_i v^c_n$, where
$\delta$ is the overlapping length and $v^c_n$  is the relative
normal velocity. The attached  disks are not allowed to rotate and 
their horizontal velocity is set to the velocity of the 
plates.

Each contact contributes to a direct force $\vec f^c$ and a torque
$\tau^c=R_i \vec f^c \cdot \vec t$ in the equation of motions.
The model does not include gravitational forces, but a viscous
force $\vec f^v_i=\gamma m_i \vec v_i$ and  a torque 
$\tau=\gamma m_i R^2_i \omega_i$ is included for each particle.
This viscous forces allow relaxation of particles without contacts.
We use the Verlet method for solving the equations of motion
\cite{allen87}.

The efficiency of the simulation is mainly determined 
by the method of contact detection. Our method searches in each step
the contacts in a list of {\it neighbours} that is called a
Verlet List. This list is constructed by taking the pair particles
whose distance of their center of mass satisfies the constraint 
$r^{ij} < R_i+R_j+\delta$. The Verlet List is updated when the maximal
displacement of the particles since the last update is larger
than $\delta/2$.  The parameter $\delta$ is chosen by making
a compromise between the storage (size of the Vertex List) and 
the compute time (frequency of list updates).  
A  Linked Cell algorithm is used to allow a  rapid calculation of 
the new Verlet List \cite{poeschel04}.

The material constants of the model are the   normal stiffness  
$k_n=1$; the tangential stiffness  $k_t=0.1$;
the normal $\gamma_n = 0.001$ and tangential  $\gamma_n = 0.0001$ 
damping frequency and the body viscosity $\gamma = 0.0001$.
The density of the disks $\rho=1$;  and the  friction coefficient
whose default value is $\mu=0.5$.  The control parameters are the applied 
pressure  $p_0=0.001$, and the velocity of the plates $v_0=10^{-6}$.  
The time step is $\Delta t = 0.05$. The most relevant parameter of this 
model is the ratio between the shear velocity $v_0$ and the velocity of 
compressional waves, which in or model is $v_p\sim R_0\sqrt{k_n/m}$. 
In our simulation $v_p\sim 1$ so that $v_p/v_0 \sim 10^6$.  This value 
should  be  compared to the ratio in realistic fault zones, where
$v_p \sim 1Km/s $ and $v_0 \sim 1 cm/year$ leading to a factor of 
$v_p/v_0 \sim 10^{10}$. To remedy this discrepancy of time scales 
we use the quasistatic limit: The velocity is chosen low enough so 
that the reduction of it by half affects the effective friction  
coefficient by less than $5\%$.

\section{Effect of particle rotation}
\label{rotation}

Here we  address the question of how particle rotation affects 
the mechanical response of the shear cell. Simulation of shear
cells with rotating and non-rotating disks are compared by 
calculating  the power dissipation and the stress ratio at the
limit of large shear deformation. 
Each shear cell consists of $1600$ disks with random sizes 
between $0.5R_0$  and  $1.5 R_0$.  The length of the cells is $80 R_0$.
We start from a loose packing that is compressed by  applying a constant 
pressure at the top plate. After a  short collisional  regime the sample 
reaches an equilibrium configuration  where the kinetic energy decreases 
exponentially with time. After this stage   the sample is sheared  by 
applying a horizontal velocity $v_0$ at the top  plate and $-v_0$ at the  
bottom. The simulations are performed by taking microscopic friction 
coefficients between  $0.0001$ and $8$. These  values should be compared 
with the friction of realistic materials,  which ranges from 
$0.0001$ for  super-lubricated surfaces to $1.2$ for rubber-concrete contact  
surfaces.
 
\begin{figure}[t]
  \begin{center}
    \epsfig{file=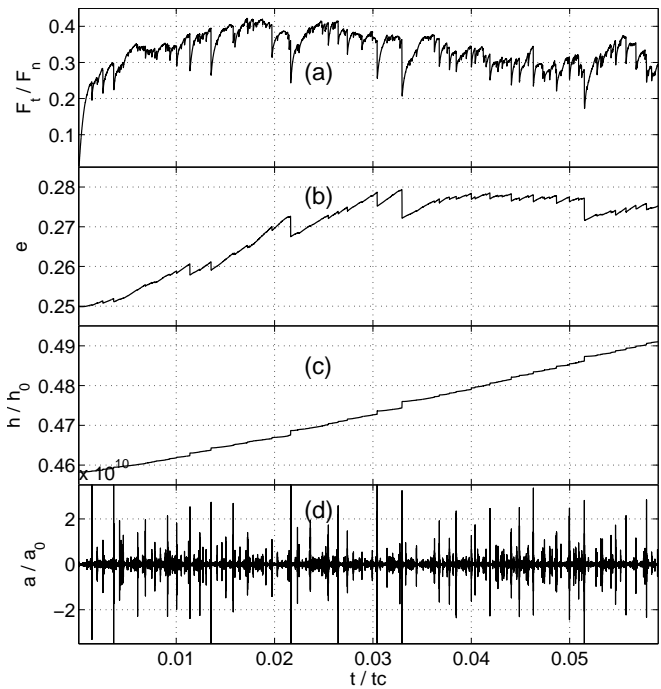,width=\linewidth}
    
   \caption{Time profiles of (a) ratio between the normal and tangential force
acting on the plates, (b) void ratio, calculated as $e=(A-A_d)/A_d$, where
$A$ is the area of the cell and $A_d$ is the total area occupied by disks,
(c) frictional dissipation normalized by $h_0= L p_0 v_0 t_c$ and
(d) acceleration of one disk in the center of the cell, normalized by
$a_0 = L/t_c^2$.  The time is normalized by $t_c = L/v_0$}
   \label{fig:test}
  \end{center}
\end{figure}

All samples reach a limit state for large shear deformation. This state
resembles the so-called critical state of soil mechanics \cite{wood90}.
As shown the Fig.~\ref{fig:test}, the stress and the void ratio reach a 
constant value besides some fluctuations. These fluctuations result 
from the characteristic stick-slip dynamics of the shear cell:  
In the stick phase, the elastic energy is stored in force chains. 
These chains build up during the slow  relative  motion of the plates. 
The elastic energy  is  released in form of {\it quakes}. Each quake  
corresponds to the collapse of a  force  chain  which is reflected by an 
abrupt drop of the macroscopic stress and a sudden compaction of the sample. 
The collapse of force chains leads to  reorganization of the particles and 
hence acoustic emission. This is  detected from the acceleration of a single 
particle  in the sample,  see part (d) of Fig.~\ref{fig:test}.  Between  two 
quakes the  buckling of  forces chains leads to an  overall  elastoplastic  
response  with  constant  increase of void ratio.

\begin{figure}[b]
  \begin{center}
    \epsfig{file=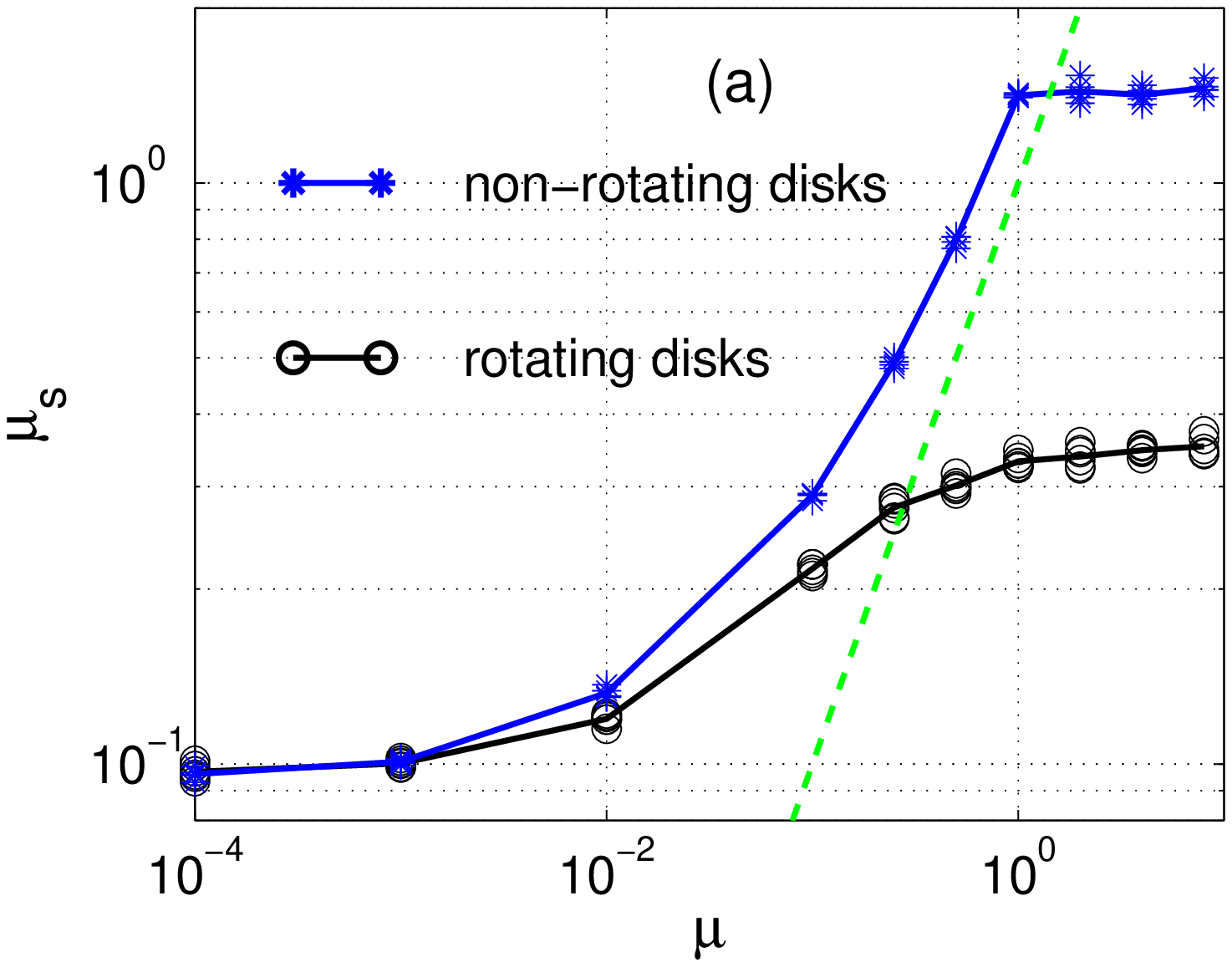,width=\linewidth}
    \epsfig{file=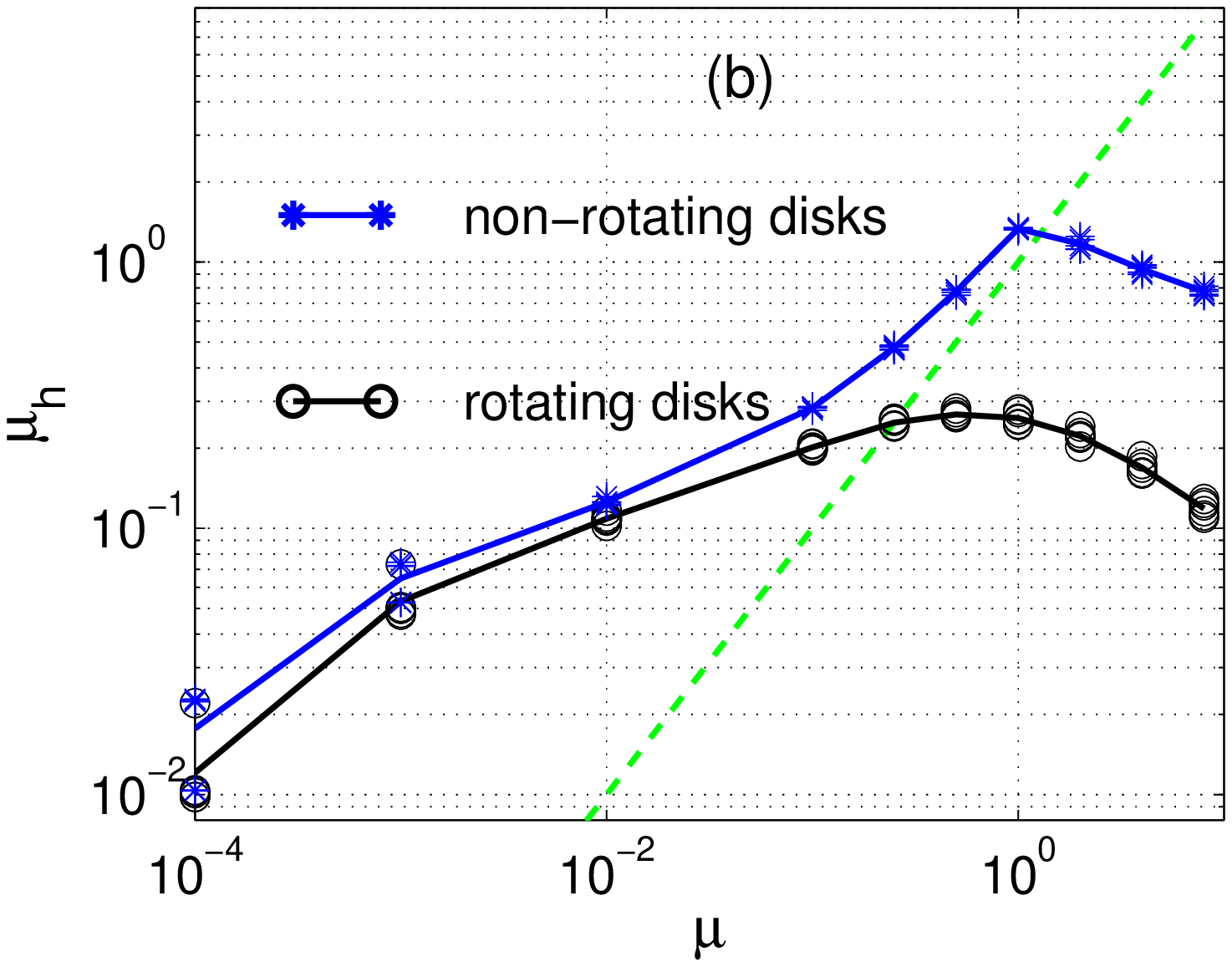,width=\linewidth}
   \caption{Dependence of  the (a) effective friction coefficient and 
            (b) frictional power coefficient on the  microscopic
            friction coefficient for rotating and non-rotating particles.
            Data of five different samples are superposed. Solid lines 
            represent the linear interpolation around the averaged value.
            dashed line corresponds to $\mu_s=\mu$ and $\mu_h=\mu$.}
   \label{fig:mu_mu}
  \end{center}
\end{figure}

The strength of the shear cell in the critical state can be quantified
by an effective friction coefficient  $\mu^*=F_t/F_n$. 
Here $F_n$  and $F_t$ are the normal and tangential force acting on the 
top plate. Part (a) of Fig.~\ref{fig:test} shows the time evolution
of  $\mu^*$. Typically this value increases in the interval  between two 
quakes,  and it drops during the  quakes, leading to time fluctuations 
which are of the same order as the time average value $\mu_s=\bar{\mu^*}$.

The dependence of $\mu_s$ on the microscopic friction coefficient
for rotating and non-rotating disks is plotted in part (a) of Fig.
\ref{fig:mu_mu}.  For small values of $\mu$ the strength of the 
shear cells is not affected by grain rotation. In both cases 
the effective friction coefficient is larger than $\mu$. This
is due to the addition of an interlocking strength component to the 
shearing friction component. In the limit case $\mu\rightarrow 0$
the effective friction tends to $0.1$, showing that interparticle
friction is not the only origin of the macroscopic frictional behavior
of granular materials.
 
For large values of $\mu$ particle rotation has a significant effects 
on the strength of the shear cell. Particularly, in the
range $0.28<\mu<1.42$  samples with rotating disks have a macroscopic 
friction coefficient lower than the contact friction coefficient. On the
other hand, the macroscopic friction coefficient is larger than the
contact friction coefficient for non-rotating disks.
In the limit  $\mu\rightarrow \infty$, the strength 
of both samples tends to a  constant value, This value is $0.34$ for 
rotating disks samples and $1.43$ for non-rotating disks.  

\begin{figure}[t]
  \begin{center}
    \epsfig{file=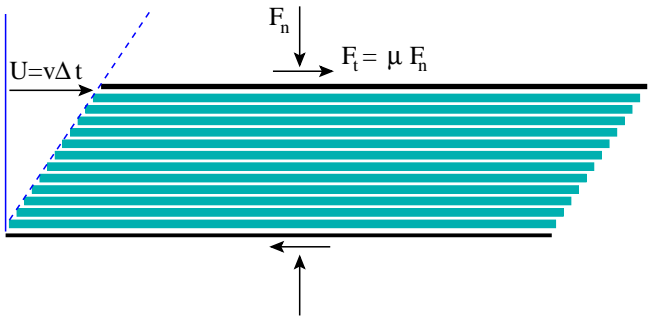,width=\linewidth}
   \caption{Simple shear model of gouge deformation.}
   \label{fig:simple_shear}
  \end{center}
\end{figure}

We now turn to the effect of grain rotation on dissipation. Energy
loss in the sample results from  frictional and viscous dissipation.
In our simulations, viscous dissipation takes place only in the 
collisional regime, which applies only during the phase of
sample preparation. In the quasistatic phase and as the sample reaches 
the critical state,  the frictional dissipation is at least  three orders 
of magnitude larger than the viscous dissipation. Therefore 
the produced heat can be calculated as:

\begin{equation}
h=\int{P(t)dt},
\end{equation}

\noindent
where $P$ is frictional power dissipation. This quantity is calculated as the 
sum of the tangential force times the sliding velocity of all the 
contacts. 

\begin{equation}
P(t)=\sum_c {f^c_t (v^c_t-\frac{d (\Delta x^c_t)}{dt})},
\label{eq:power}
\end{equation}

\noindent
where $f^c_t$ is the tangential contact force;  $v^c_t$  the tangential 
relative velocity at the contact and $\Delta x^c_t$ the elastic part of the 
tangential displacement as defined by Eq. (\ref{eq:dxt}).  The heat versus 
time is plotted in part (c) of  Fig.~\ref{fig:test}  for rotating 
particles with $\mu=0.5$.  The general trend of the heat is to increase 
slowly between two quakes, and rapidly during quakes. 

The calculated heat should be compared to the theoretical energy 
dissipation of simple shear,  which reads:
 
\begin{equation}
h_{th} = \mu F_n v \Delta t,
\label{eq:heat_th}
\end{equation}

\noindent
where  $\mu$ is the microscopic friction coefficient, $F_n$  the 
normal force on the plates, $v=2v_0$ the relative velocity of the 
plates and $\Delta t$ is the time interval.  
The  value of Eq.~(\ref{eq:heat_th}) is 
calculated according to the model shown in Fig.~\ref{fig:simple_shear}.
This model corresponds to the simple shear of a multilaminate 
with interlaminate friction equal to $\mu$. To compare this value to 
the calculated power dissipation $P(t)$, we introduce the frictional power 
coefficient as  $\mu_h = \overline{P(t)}/P_0$, where   
$P_0=F_n v$ and  $\overline{P(t)}$ is the time average along the critical
state regime. Note that  $\mu_h=\mu$ when the measured heat coincides 
with the theoretical value. 

Part (b) of Fig \ref{fig:mu_mu}  shows the dependency of the frictional
power coefficient  on $\mu$. Heat production assumes a maximal 
value at $\mu=0.5$ for rotating particles and $\mu=1.0$ for non-rotating 
particles. In both cases the dissipation tends to decrease for 
$\mu\rightarrow 0$ and  $\mu\rightarrow\infty$.  In the first case because 
$f_t\rightarrow 0$ and in the second  case because the fraction of sliding 
contacts is very small for large microscopic friction coefficients. 

For small values of $\mu$ heat production is not affected by particle 
rotation, but for larger values it is affected by almost one order of 
magnitude.
In both cases heat production is quantitatively different from the value of
simple shear. Small friction coefficients lead to a dissipation 
larger than the theoretical one. This is due to the large population
of sliding contacts in this regime, and their relative orientation with
respect to the principal direction of load, which leads to tangential 
forces larger than the expected value for simple shear. For large friction 
coefficients few contacts can reach the sliding condition, so that 
the dissipation is much  lower  than that one expected for 
simple shear. 


\begin{figure}[b]
  \begin{center}
    \epsfig{file=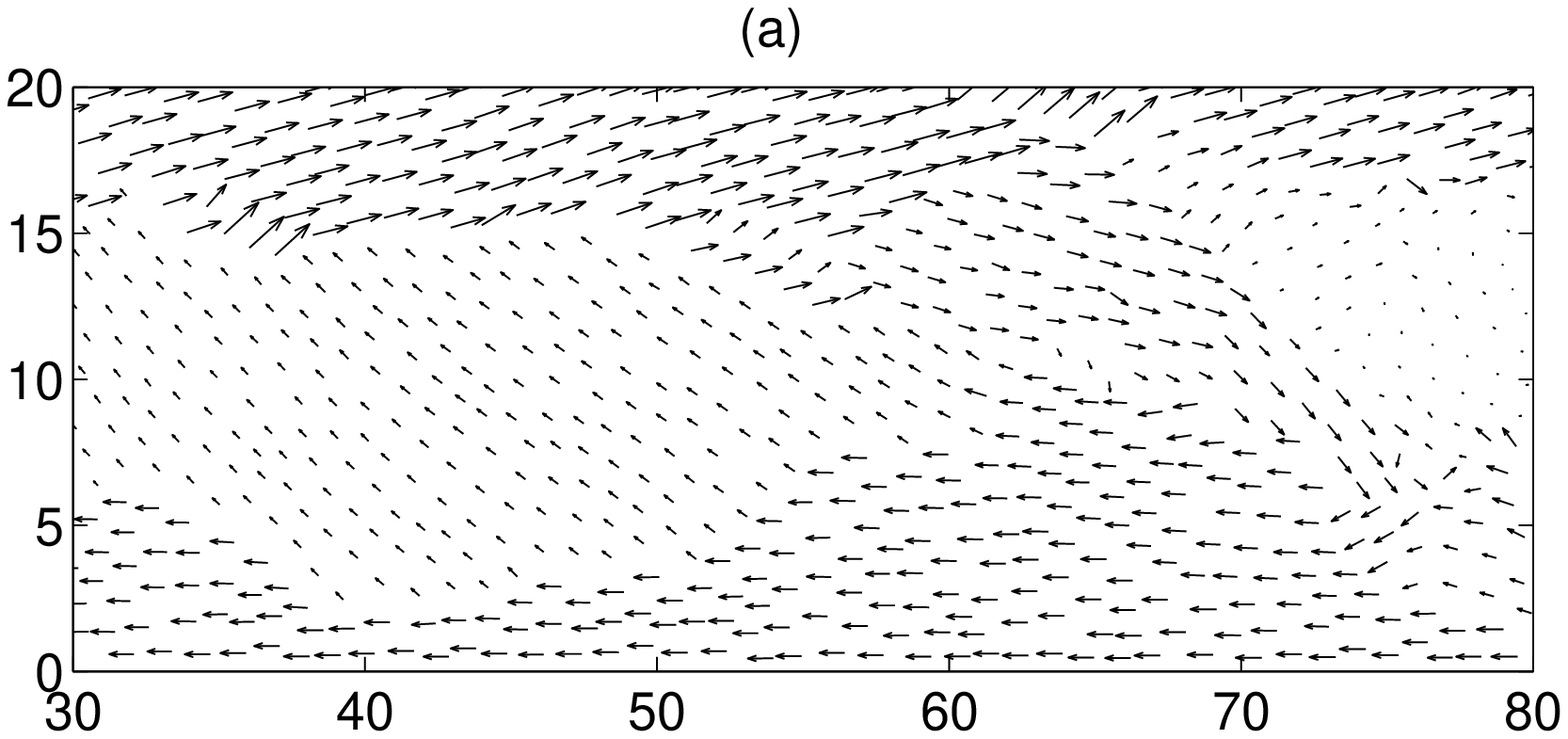,width=\linewidth}
    \epsfig{file=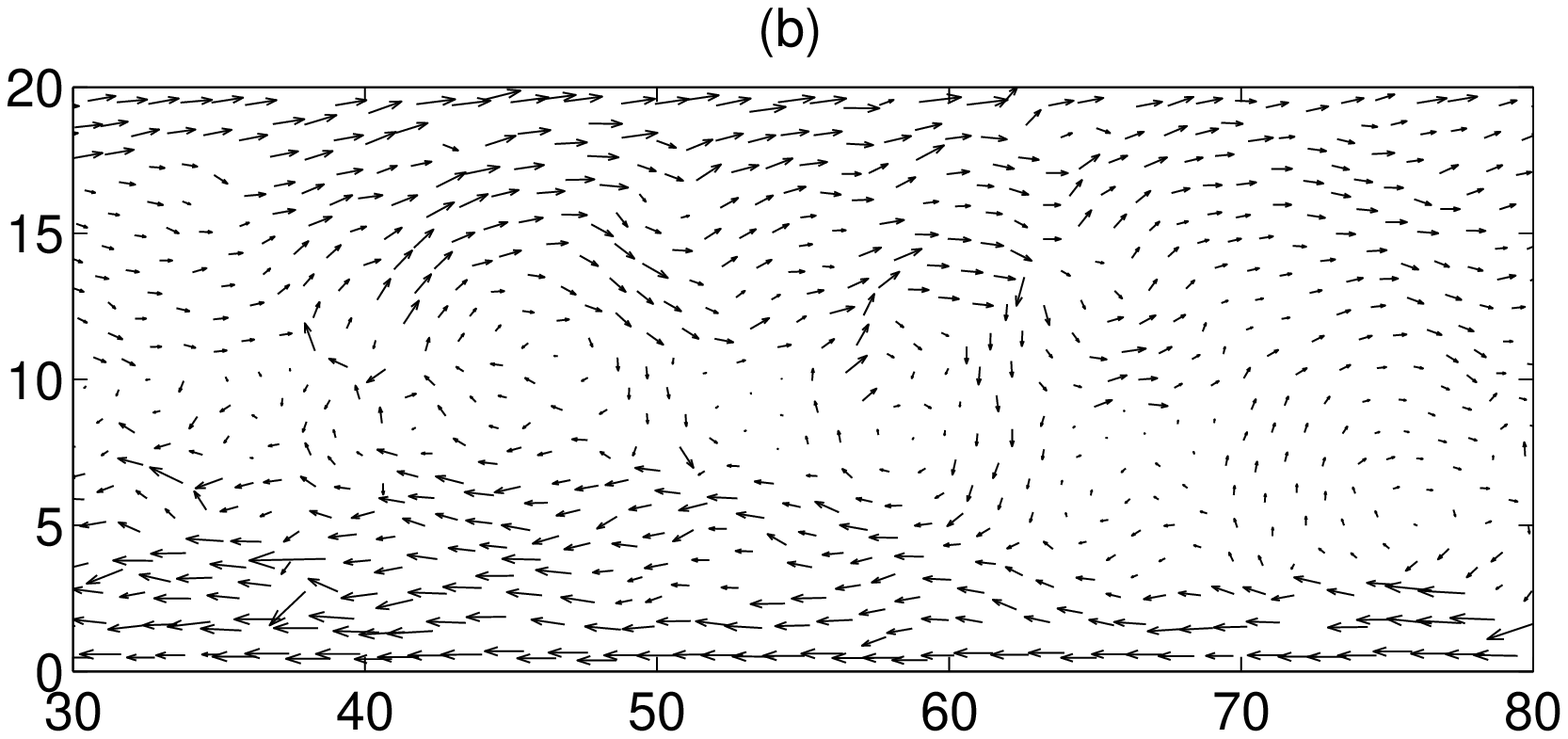,width=\linewidth}
   \caption{Snapshot of the displacement field for (a) non-rotating and 
   (b) rotating grains. The  friction coefficient is $\mu = 0.5$.}
   \label{fig:dfield}
  \end{center}
\end{figure}

The displacement field confirms that the picture of simple
shear (Fig.~\ref{fig:simple_shear}) is also not supported.  In part (a) 
of Fig.~\ref{fig:dfield} we plot the displacement 
field for non-rotating  particles when $\mu=0.5$. 
We observe the formation of blocks  of particles moving approximately as 
a whole. The boundaries between  these blocks are not flat, but curved. 
This produces strong interlocking, that explains the  considerable 
increase of the strength of the material with respect to simple 
shear. Note that these blocks can not slide against each other for very 
long  without changing shape. The reorganization of these structures appears
by means of large quakes. These quakes change completely the displacement
fields, even when the position of the particles stays approximately the same.

If the particles are allowed to rotate, vorticities appear  spontaneously. 
They are shown in part (b) of Fig~\ref{fig:dfield}. These vorticities
are accompanied by strong temporal fluctuations of the displacement field, 
but they can appear and disappear intermittently in the same zone. 
The vorticity field has been observed in many numerical simulations   
\cite{sakagushi97, kuhn99}. It resembles to some extent the  turbulent 
regime observed in fluids, but their dynamics is quite different
~\cite{radjai02}: Fluids under slow  motion present a  laminar regime where 
the  mass displacements can be  considered as simple shear.
On the other hand, our shear cells develop large vorticities even in the 
quasi-static regime, ruling out the simple shear deformation regime. The 
short lifetime of these   vorticities, typically the same as the interval
between two large quakes,  contrasts to the large lifetime of the eddies in 
turbulent flow.  Thermal  measurements in dynamic shear-banding on metals 
also reveals that homogeneous shear is not possible and vortices are the 
rule \cite{guduru01}.

The spatial distribution of vorticity cells, when combined with the 
distribution of rolling, provides a novel picture of bearings:
Inside vorticity cells all particles rotate as a rigid body,
whereas the space between the vorticity cells is characterized by intense 
relative rotations. This leads to clusters of rotational bearings and zones  
of  intense slippage. 
We will quantify the contribution of these modes to 
global deformation by performing a kinematic decomposition  of the 
contact deformation.  This  decomposition will distinguish  
rolling from sliding and from the  rigid body motion of the  vorticity cells.

\section{Measure of rolling}
\label{rolling}

\begin{figure}[t]
  \begin{center}
    \epsfig{file=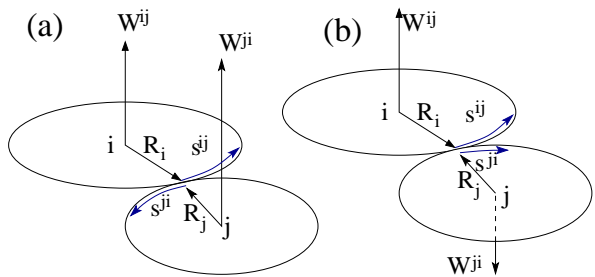,width=0.9\linewidth}
  
   \caption{homothetic (a) and antithetic (b) couple of the Cosserat rotations. }
   \label{fig:cosserat}
  \end{center}
\end{figure}

Several definitions of rolling can be found in the literature
\cite{shodja03,kuhn04,bagi04}. They are taken as measures of 
contact  deformation for each pair of interacting particles.  Kuhn and 
Bagi decompose the degrees or freedom of the two  particles into rigid 
body  motion and objective motion \cite{kuhn04}. 
Then the rolling is defined as the average of the objective translation 
at each side of the contact point \cite{bagi04}. Based on these studies, we 
introduce a new definition of rolling in terms of the rotation and
translation of the particles in contact. This definition will be consistent 
with the fact that rolling allows deformation in the assembly without 
frictional dissipation or accumulation of elastic energy. 

Let us suppose that at time $t$ two particles indexed by $i$ and $j$ are
in contact, and they stay in contact during an infinitesimal time 
interval  $dt$ afterward. We introduce a  system of coordinates 
${\vec n^{ij}, \vec t^{ij}, \vec z^{ij}}$ attached to the contact. 
The unit normal vector $\vec n^{ij}$ connects the center of mass of the 
particle  $i$ to that of the particle $j$:

\begin{equation}
\label{eq:n}
\vec n^{ij}=\frac{\vec x_j(t)-\vec x_i(t)}{|\vec x_j(t)-\vec x_i(t)|} \approx 
        \frac{\vec x_j(t)-\vec x_i(t)}{R_i+R_j}.
\end{equation}

\noindent
The latter approximation is valid when $\Delta x_n << d$, where  $\Delta x_n$
is the overlapping length and $d$ the characteristic diameter of the disks.
The unit vector $\vec z^{ij}$ is perpendicular to the plane of the disks, and
the unit tangential vector satisfies 
$\vec t^{ij} = \vec z^{ij}\times\vec n^{ij}$.

Let us consider two points attached to each particle, in a region 
infinitesimally near to the contact point.  The tangential 
velocity of these points   are given in terms  of the linear 
$\vec v_k$ and angular $\omega_k$  velocities of the particles:

\begin{eqnarray}
\label{eq:vc}
v^c_{i,t} &=& \vec v_i\cdot \vec t +  \omega_i R_i\nonumber\\
v^c_{j,t} &=& \vec v_j\cdot \vec t -  \omega_j R_j. 
\end{eqnarray}

Let us take the movement of the  branch vector. i.e the
vector connecting the center of mass of the two particles.
The tangential component of the velocity of a point attached
to the branch vector at the contact  will be called 
{\it rigid body velocity}:

\begin{equation}
\label{eq:rb}
V^{ij}_{rb}=\frac{(\vec v_i R_j + \vec v_j R_i)\cdot \vec t}{R_i+R_j}.
\end{equation}

\noindent
If the two particles move as a rigid body, the velocity coincides 
with Eqs. (\ref{eq:vc}). Otherwise there is a relative velocity between the
two points attached to the particles.  This velocity can be calculated 
by subtracting the rigid body like velocity of Eq. (\ref{eq:rb}) from 
the contact velocities of Eqs. (\ref{eq:vc}):

\begin{eqnarray} 
s^{ij} &=&  v^c_{i,t}- V^{ij}_{rb} \nonumber\\
s^{ji} &=&  v^c_{j,t}- V^{ij}_{rb}
\end{eqnarray}

\noindent
These velocities can be given in terms of linear and angular velocities 
by using  Eqs.~(\ref{eq:vc}) and (\ref{eq:rb}):

\begin{eqnarray}
\label{eq:obj}
 s^{ij} &=&   \omega_i R_i - 
                \frac{R_i}{R_i+R_j}(\vec v_j-\vec v_i)\cdot \vec t^{ij},
\nonumber \\
 s^{ji} &=&  -\omega_j R_j + 
                \frac{R_j}{R_i+R_j}(\vec v_j-\vec v_i)\cdot \vec t^{ij}.
\end{eqnarray}

\noindent
They will be called {\it objective velocities} of the two particles at 
the contact. These velocities are {\it objective} in the sense that their 
magnitude is unaffected by the common rigid-body motion of the 
particle pair. In particular, these measures are not affected by 
a change of the reference frame. The objective velocities vanish if,
and only if, the two particles move as a rigid body. Otherwise
the objective velocities involve rolling, sliding or accumulation
of shear strain at the contact.

We will relate these {\it objective velocities} to the aforementioned 
Cosserat-continuum rotations. At the micromechanical level, the later 
are defined as  the  relative rotation of the  particle  with respect 
to the rotation of the  branch vector around the axis parallel to 
$\vec z^{ij}$ at the contact point ~\cite{vardoulakis95}:
 
\begin{eqnarray}
W^{ij} & = &\omega_i-\Omega^{ij},\nonumber\\
W^{ji} & = &\omega_j-\Omega^{ij}
\label{eq:cosserat}
\end{eqnarray}

\noindent
where $\phi_i$ is the angular velocity of the particle $i$ with radius
$R_i$. $\Omega^{ij}$ is the rotational velocity of the branch vector:

\begin{equation}
\label{eq:omega}
\Omega^{ij} = \frac{(\vec v_i - \vec v_j)\cdot\vec t^{ij}}{R_i+R_j}
\end{equation}
  
Replacing Eq.~(\ref{eq:omega}) into Eqs.~(\ref{eq:cosserat}) and comparing 
the result with Eqs.~(\ref{eq:obj}) we obtain a relation between 
the Cosserat rotations and the objective velocities.

\begin{equation}
\label{eq:theorem}
  s^{ij}=  R_i W^{ij}~~~~~~s^{ji}= - R_j W^{ji}
\end{equation}

These equations provide an interesting connection between the relative
orientation of the Cosserat rotation and rolling: 
When the Cosserat rotations are in a {\it homothetic couple} 
as shown part (a) of Fig.~\ref{fig:cosserat},  the objective 
velocities  $s^{ij}$ and  $s^{ji}$ have opposite signs.
In this case, depending on whether the contact is or not in the 
sliding condition, the contact  deformation results either in 
frictional dissipation or in accumulation of elastic energy,
without rolling deformation. Homothetic couples appear in 
several cases: If the two disks in contact rotate 
in the same sense without linear velocity, the Cosserat rotations 
reads $W^{ij}=\omega_i$ and  $W^{ji}=\omega_j$.  If the 
particles do not rotate, the Cosserat rotations are given by 
$W^{ij}=W^{ji}=\Omega^{ij}$. In both cases they are in 
a homothetic couple, and hence there is no rolling deformation.

\begin{figure}[t]
  \begin{center}
    \epsfig{file=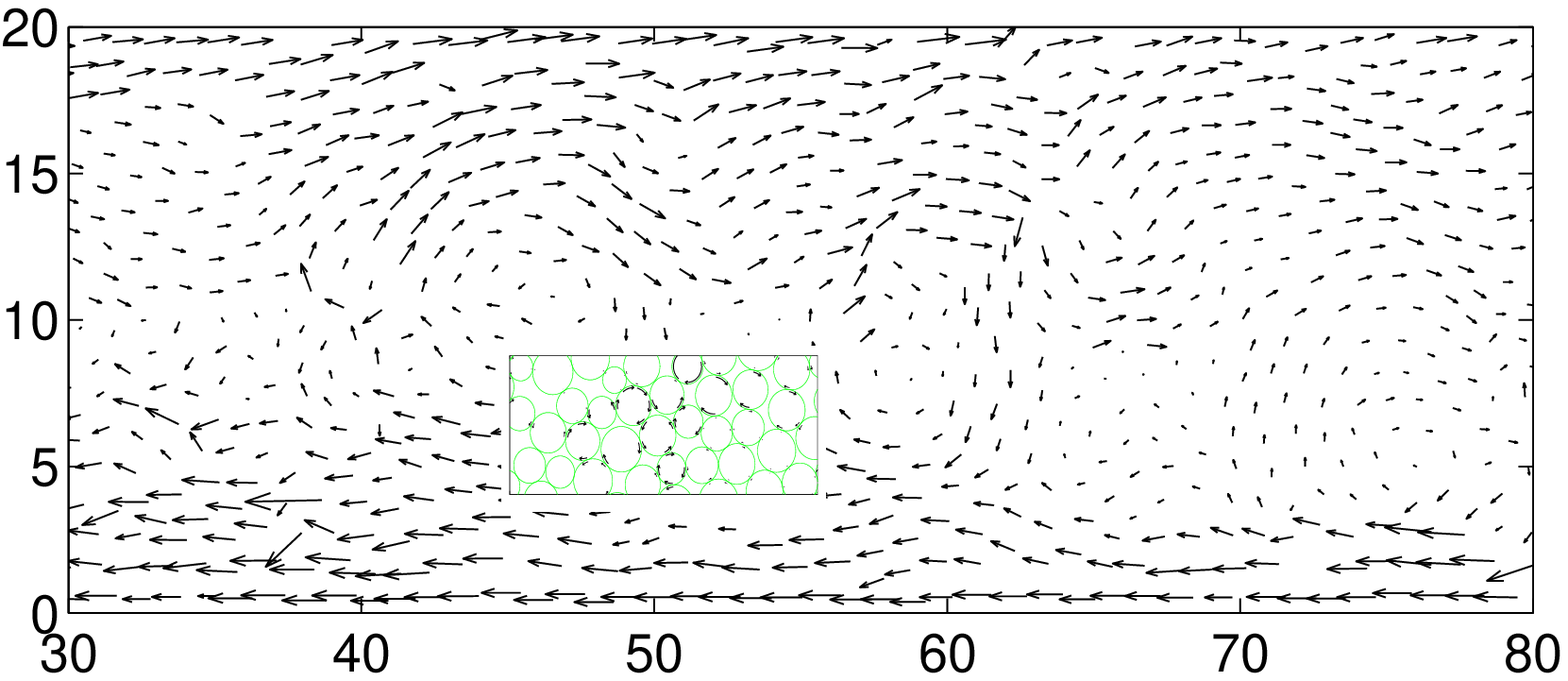,width=\linewidth}
    \epsfig{file=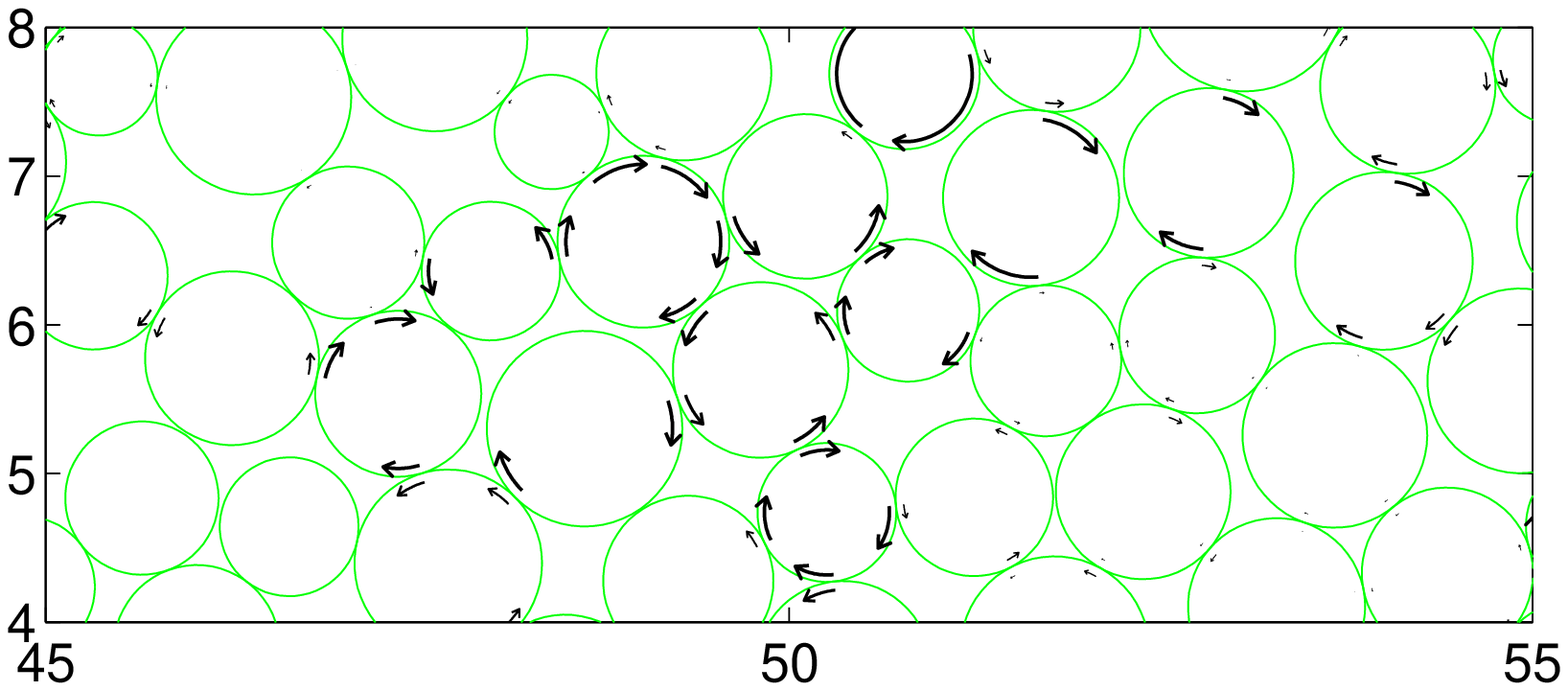,width=\linewidth} 
    \epsfig{file=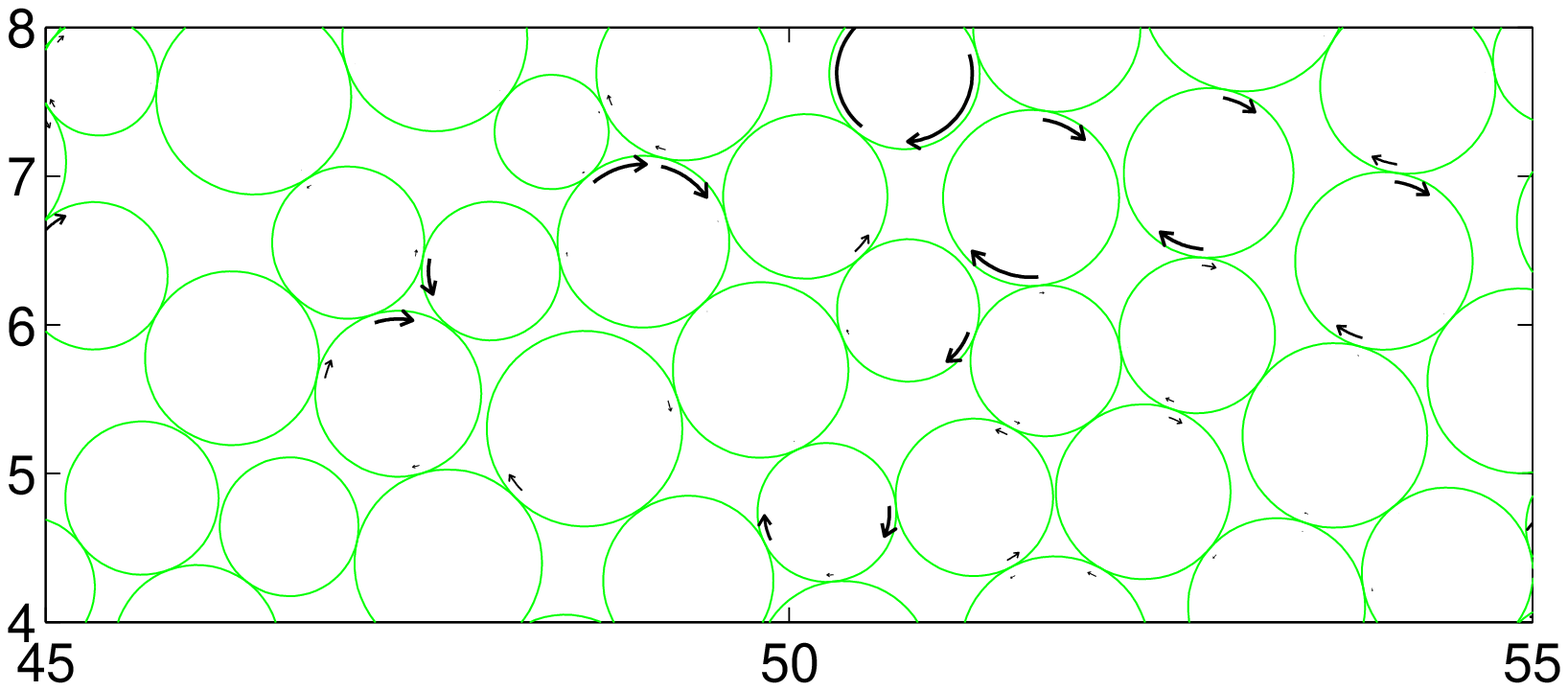,width=\linewidth}
    \epsfig{file=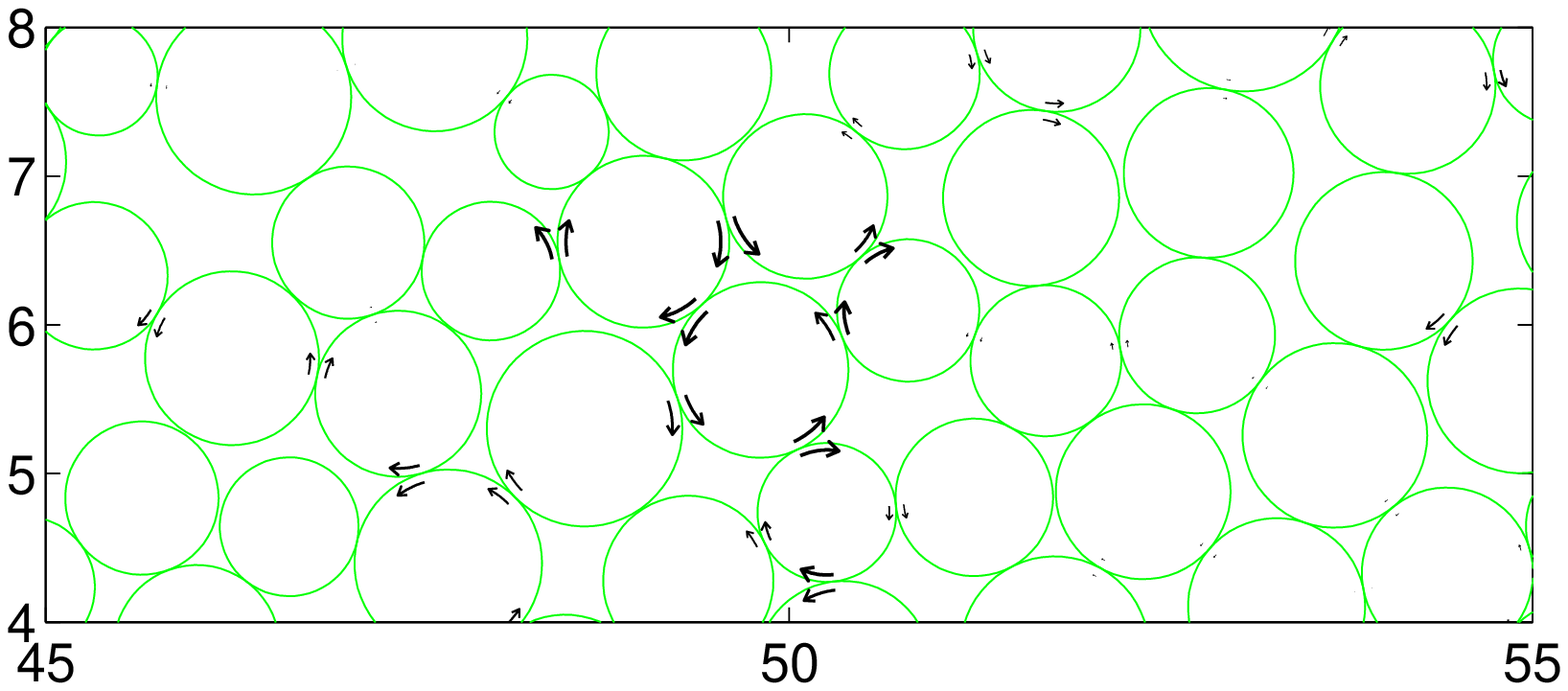,width=\linewidth}
   \caption{Snapshot of (a) vorticity field, (b) objective velocities at 
the contacts   as shown the inset in part (a), and their decomposition in  
(c) dislocation and (d) rolling. The sample consists of rotating grains
with $\mu=0.5$. }
   \label{fig:slid_roll}
  \end{center}
\end{figure}

  
The appearance of rolling between the particles is linked to the
{\it antithetic couple} of the Cosserat rotations. This couple is
shown in part (b) of Fig.~\ref{fig:cosserat}. In this case the
rolling is given by the common part of the objective velocities,  
i. e. $\min(|s^{ij}|,|s^{ji}|)$. In particular, if 
the Cosserat rotations satisfy $W_i R_i+W_j R_j=0$ both grains
have the same objective velocity. Therefore the contact 
deformation does not accumulate elastic deformation or produce frictional 
dissipation, and hence the deformation corresponds to pure rolling. 

In the general case  the objective velocities should be decomposed into 
rolling and dislocation. The  first one results in:

\begin{equation}
   V^{ij}_{roll} = \left\{
              \begin{array}{ll}
                 0 & homothetic~ couple\\
                 \min(|s^{ij}|,|s^{ji}|) sign (s^{ij}) & antithetic~ couple,
               \end{array}
                      \right.
\end{equation}

\noindent
and the dislocation is given by the difference  of the objective velocities:

\begin{equation}
\label{eq:vdisl}
V^{ij}_{dis} = s^{ij}- s^{ji}.
\end{equation}

\noindent
Using the identity $s^{ij}-s^{ji}=v^c_{i,t}-v^c_{j,t}$, 
it is easy to prove that the dislocation velocity corresponds to the  
relative   tangential  velocity  defined in Eq. (\ref{eq:reltanvel}). 
In the limit of rigid disks, where elastic deformation at the 
contact is absent, the dislocation velocity coincides with the sliding 
velocity. In granular dynamics an elastic shear deformation is allowed, so
that the sliding velocity is given by:

\begin{equation}
\label{eq:disl3}
V^{ij}_{slid} = V^{ij}_{disl}-\frac{d (\Delta x_t)}{dt},
\end{equation}

\noindent
where $\Delta x_t$ is the elastic part of the tangential displacement
at the contact, that is given by Eq. (\ref{eq:dxt}). According to 
Eq.~(\ref{eq:power}) the sliding velocity times the tangential
force corresponds to the frictional dissipation. The second term 
of Eq.~\ref{eq:disl3} involves accumulation of elastic energy.

The Cosserat rotations turn out to be a suitable {\it order parameter} 
to describe the vorticities and bearings in the shear deformation: 
As shown the Fig.~\ref{fig:slid_roll}, the spatial distribution of these 
structures is not just random, but strong correlations appear in form of 
three  coexisting  phases: (1) Vorticity cells, where the 
particles rotate almost as a whole, so that the Cosserat rotation is
vanishingly small, (2) clusters of particles with intense rolling 
(rotational bearings), where the Cosserat rotations are antithetic,
and (3) zones between particles with intense dislocation (microbands),
characterized by homothetic couples of the Cosserat rotations.
Microbands induce frictional  dissipation and accumulation of elastic 
energy, whereas the rotational bearings   accommodate the vorticity cells 
to make them more compatible with the imposed  kinematic boundary 
conditions.

\begin{figure}[t]
  \begin{center}
    \epsfig{file=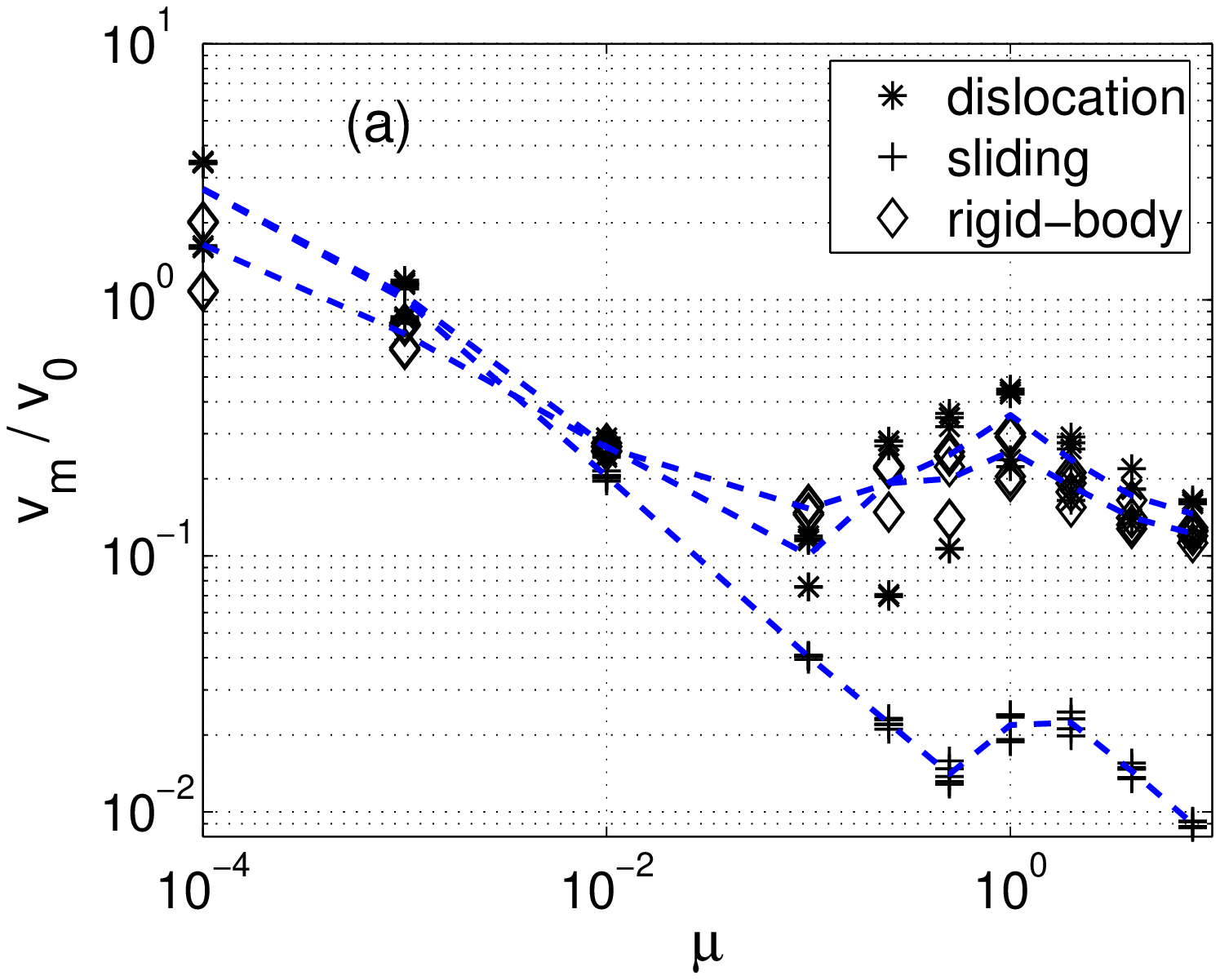,width=\linewidth}
    \epsfig{file=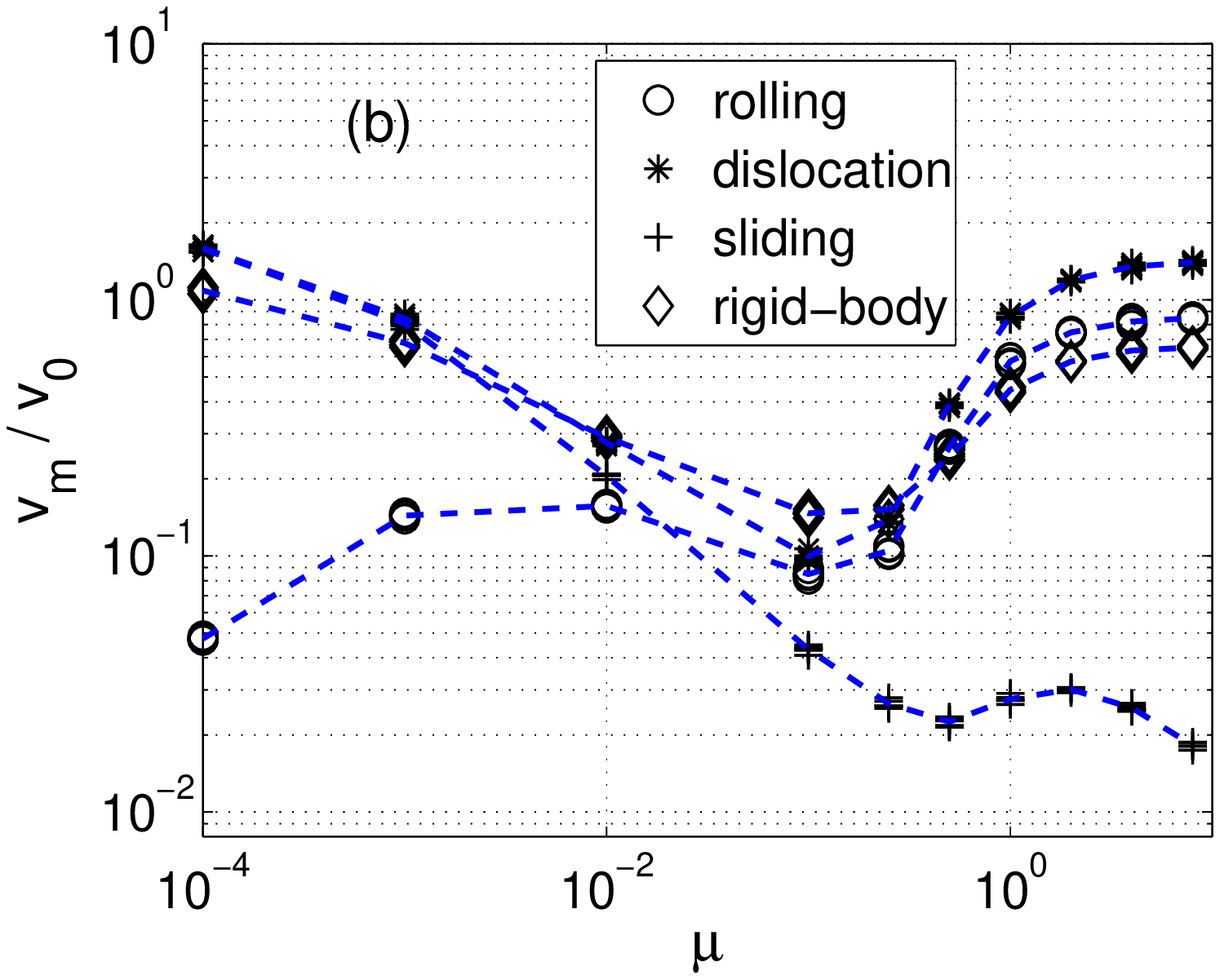,width=\linewidth}
   \caption{Decomposition of kinematic roles for (a) non-rotating
            and (b) rotating particles, as function of $\mu$. 
            They are calculated form the average of the absolute value of 
            the rolling, dislocation, sliding and rigid body velocities at the
            contacts. Data of five different samples are superposed.}
   \label{fig:modes_mu}
  \end{center}
\end{figure} 

The microscopic friction coefficient can serve as a control parameter
of the relative  population of these three phases. This is shown
in Fig.~\ref{fig:modes_mu} for (a) non-rotating and (b) rotating
particles. In both cases the sliding displacement and rigid body 
velocities are the dominant deformation modes for small values of 
$\mu$. In this regime, the dislocation velocity coincides with the sliding 
velocity, because almost all the contacts are in the sliding condition.  
Deformation in this case is characterized by small clusters of particles  
moving against each other through microbands of intense sliding.  For 
rotating  particles the rolling is much lower than the other deformation 
modes, whereas the rolling is absent for non-rotating grains. 

For large friction coefficients the sliding velocity becomes negligible,
because few contacts are able to reach the sliding condition. In
this regime, depending on whether the particles are or not allowed to
rotate, the population of the deformation modes is quite different:  
for non-rotating particles, the contact deformation is dominated by 
rigid-body motion and elastic dislocation. The latter one builds up 
elastic energy that is liberated in the form of  strong quakes. 
The dependence of dislocation and rigid-body motion on $\mu$ is fairly
weak.  This reflects a self organization of  the shear  cells, characterized 
by a non-dependency of the effective friction coefficient on $\mu$, 
already shown  in part (a) of Fig.~\ref{fig:mu_mu}.

For rotating grains and large friction coefficients rolling plays a 
relevant role, as shown part (b) of Fig.~\ref{fig:modes_mu}.
The deformation is dominated by rigid-body motion  (due to the vorticities),  
rolling (due to the rotational bearings) and  elastic dislocation
(due to building of force chains). The self organization of the shear cells 
is given by the fact that  rolling and vorticities are not affected by a 
change in the microscopic  friction coefficient. 
The  sliding  turns out to be much  smaller that the 
other modes, because only few contacts can reach the sliding condition. 
Therefore the   dislocation results almost completely in accumulation of 
elastic energy  that is released in the form of quakes.

\begin{figure}[t]
  \begin{center}
    \epsfig{file=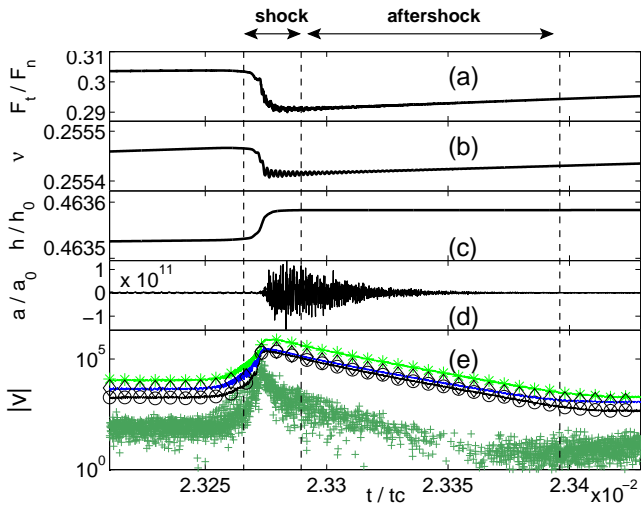,width=\linewidth}
   \caption{Time evolution of (a) stress, (b) void ratio,            
   (c) frictional dissipation, and (d) acoustic emission at the center of 
   the sample.(e) Average of the absolute value of rolling (circles), 
   dislocation (asterisks), sliding (crosses) and rigid body velocities 
   (diamonds). The sample consists of rotating particles with $\mu=0.5$.}
   \label{fig:quake}
  \end{center}
\end{figure}

The details of the dynamics of one quake is shown in Fig.~\ref{fig:quake}.
We use a friction coefficient of $\mu=0.5$. We observe a small time interval 
(shock) characterized by a sudden decrease 
of the effective friction coefficient, along with an abrupt compaction 
and a rapid generation of heat. This is followed by a longer time interval 
(aftershock) given by an exponentially decay of microseismic activity with
almost no frictional dissipation. 
More details of the structure of the quake are visible by calculating the
time evolution of the contact modes of deformation. Part (e) of
Fig.~\ref{fig:quake} shows that most of the contact deformation at the shock 
corresponds to rolling and rigid body deformation. In the 
after-shock the sliding velocities become  vanishing small, whereas the 
other modes decay exponentially. Acoustic waves are generated at the point 
where the force chain fails. These waves travel though the sample and they 
are reflected as they reach the plates. After many reflections they  becomes 
uncorrelated, which removes almost all contacts from the sliding condition.  
Therefore sliding  velocity modes are not active in the aftershock, whereas 
the other modes decay  exponentially in form of uncorrelated oscillations. 

As far as earthquakes is concerned, it is remarkable that only a minute part 
of the contact deformation during the quake corresponds to sliding.
This leads to an effective friction coefficient around $0.25$ which is  
lower than the contact friction coefficient of $\mu=0.5$, see part (b) of
Fig.~\ref{fig:mu_mu}.  Therefore the introduction of rolling in the gouge 
dynamics could potentially  explain the Heat Flow  Paradox. 

\section{Concluding remarks}
\label{conclusion}

The most important contribution of this work is to show 
that spontaneous formation of vorticity cells assisted by 
rotational bearings is the mechanism of reduction of 
strength and frictional dissipation in shear cells. For
the range of rock friction of $\mu=0.6-0.9$, the
effective friction coefficient increases from $\mu_s=0.3$
to $\mu_s=0.35$, whereas the frictional power coefficient 
decreases from $\mu_h=0.28$ to $\mu_h=0.22$. These results
could explain the heat flow anomalies in fault gouges,
where the macroscopic friction  coefficient,  as deduced from 
heat flow observations, is around one  order of  magnitude 
lower than the rock friction \cite{lachenbruch92}.
The existence of such rotational patterns confirm  earlier 
speculations about the effect of rolling  in the  reduction of 
heat production by means of a self-organization process in 
fault gouge \cite{mora99}. 

The effective friction coefficient and frictional power coefficient are 
strongly dependent on particle rotation. This reflects the necessity to 
introduce rotational degrees of freedom in the continuum 
description of fault gouge. As deduced from the deformation field at the 
contacts,  the Cosserat continuum approach should be consistent with the 
observed  three phase separation of kinematic modes: 
(1) vorticity cells, where the Cosserat rotations are absent; 
(2) Bearings given by antithetic couple of the Cosserat rotations, and 
therefore a pronounced rolling at the contacts; and 
(3) microbands of homothetic couples, with pronounced dislocations, and 
hence, strong accumulation of elastic energy and high frictional dissipation.
This description requires an extension of the existing Cosserat continuum 
models, which consider only homothetic couples of the Cosserat rotations
\cite{papanastasiou92,tejchman96}. 
As far as dissipation is concerned, besides the Cauchy stress 
tensor, a couple stress tensor should be introduced as an additional static 
variable of the enhanced continuum, entering as the energetically dual 
counterpart of the gradient of the homothetic part of the Cosserat rotations 
\cite{tordesillas04,ehlers03,froiio05a,muhlhaus87b}.   
In this context numerical simulations can be used as a virtual 
laboratory to assist the development of micromechanical constitutive models.

An improved constitutive law for fault gouge will require integration of
not only rolling, but also the contribution of grain fragmentation 
to the energy budget. Observations of  fault zones suggest that earthquakes
can pulverize rocks in the gouge \cite{wilson05}.  This  leads to a fractal 
grain size distribution and  an increase of the gouge surface area. Some  
important issues concerning  the heat flow paradox should be considered: 
Does fracture energy play an important role in the earthquake energy balance? 
and do such fractal gouges develop vortical  structures and rotating bearings?.

There are also some aspects about the aseismicity of fault zones which  
deserve detailed study  in a future work.   Vortical structures and 
rotating bearings promote a coherent deformation which  remains  during 
two quakes. These may inhibit the large events we observe in samples with 
non-rotating  grains. The dynamics of these rotational patterns will be 
significant in  the understanding the enigmatic aseismic creep, where two 
tectonic plates  move against each  other without accumulating elastic 
energy  or generating  earthquakes \cite{scholz98,melbourne03}. 

\section*{Acknowledgments}
 
We thank S. Latham, H. M\"uhlhaus, A. Tordesillas, K. Bagi, H. Sakaguchi, 
Y. C. Wang and  F. Froiio and A.~A. Pena-Olarte for helpful discussions.  
This research was supported by the European DIGA project HPRN-CT-2002-00220 
and the Australian Computational Earth Systems Simulator (ACcESS). 
HJH acknowledges support from the German-Israeli Foundation grant 
No.I-795-166.01/2003. Simulations were performed on the ACcESS SGI Altix 
3700 supercomputer.


\end{document}